\documentclass[conference]{IEEEtran}
\usepackage{caption}
\usepackage{amssymb}
\usepackage{amsmath}
\usepackage{subcaption}
\usepackage{graphicx}
\usepackage{balance}
\begin{document}

\title{Autonomous line follower robot controlled by cell culture}

\author{\IEEEauthorblockN{Sayan Biswas}
\IEEEauthorblockA{Department of Electrical Engineering\\
Jadavpur University, Kolkata, India\\
Email: sayanbiswasjuee@gmail.com}}

\maketitle

\begin{abstract}
Neuro-electronic hybrid promises to bring up a
model architecture for computing. Such computing architecture
could help to bring the power of biological connection and
electronic circuits together for better computing paradigm. Such paradigms for solving real world tasks with higher accuracy is on demand now. A
robot as a autonomous system is modeled here to navigate
following a particular line. Sensory inputs from robot is directed
as input to the cell culture in response to which motor commands
are generated from the culture.
\end{abstract}
\section{Introduction}  
Computation carried by brain are fast and they are able to take rapid decisions. At times brain can take really fast decision at near zero reflex time. Although all kind of decision making does not fall into this category, but obviously there are instances where one can take decision in near zero reflex time. In a visual search task to discriminate between a car image and animal image, near zero reflex time is expected. Decision making regarding colours whether is black or whites takes almost zero reflex time. Such computation are carried by brain real fast. Such abilities of the brain have inspired researchers.  Many strategies have been developed inspired from such abilities of brain. Some of those strategies are neuromorphic devices \cite{suri2011phase} \cite{yu2011electronic} \cite{indiveri2001neuromorphic}, artificial neural network \cite{dayhoff2001artificial}, fuzzy systems \cite{nauck1997foundations} \cite{lin1996neural}. \\
An ongoing research is being carried out for making hardwares suitable to mimic neuronal systems \cite{suri2011phase}. There are challenges including versatility connectivity which limits the usage of cultures like brain. 
Neuro electronic hybrid systems can prove to be a possible computing architecture. On connecting cultures of neurons properly with the real world such abilities of the brain can be used to solve real world problems. This might prove to be a essential platform to use biological network for computing. Neurons are basic computation and functional unit of brain which are versatile in terms of behavior. The neurons in the brain form inter connected set of networks which are dynamic in nature. Such dynamic nature of neuron in network gives them learning abilities and makes them powerful enough to serve varied types of function. Hence making the brain a robust computational unit of brain. A hybrid system attempts to solve real world problems by implementing this abilities of brain.\\  MEA dish proves advantageous and promising for understanding network activity and properties of neuronal network. Robotic control through such culture \cite{bakkum2008spatio} \cite{novellino2007connecting} and training \cite{ruaro2005toward} of neuronal network is an ongoing work at various labs. There are several challenges required to be resolved for a practical and real world application of such systems \cite{george2015robot}.  Decoding output from an input to a network is open problem. To understand such decoding pattern a deeper understanding of dynamic network system \cite{biswas2016proposal} \cite{biswas2016extraction} \cite{biswas2017how} \cite{biswas2017novel} \cite{bullmore2009complex} is essential.
 \\ The description here deals with a framework for using cell cultures as a information processing tool, and to implement the ability for the purpose of classification of sensory inputs received and accordingly take decision of generating correct motor commands. The framework shall elaborately explain a proposal for implementation in real life.

\section {Line follower}

\begin{figure}[h!]
\centering
\includegraphics[width=1\linewidth]{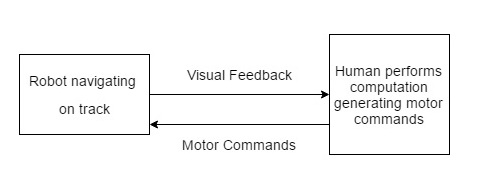}
\caption{Robot Navigation by Human}
\label{fig:navghum}
\end{figure}

\begin{figure*}
\centering
\includegraphics[width=1.25\linewidth]{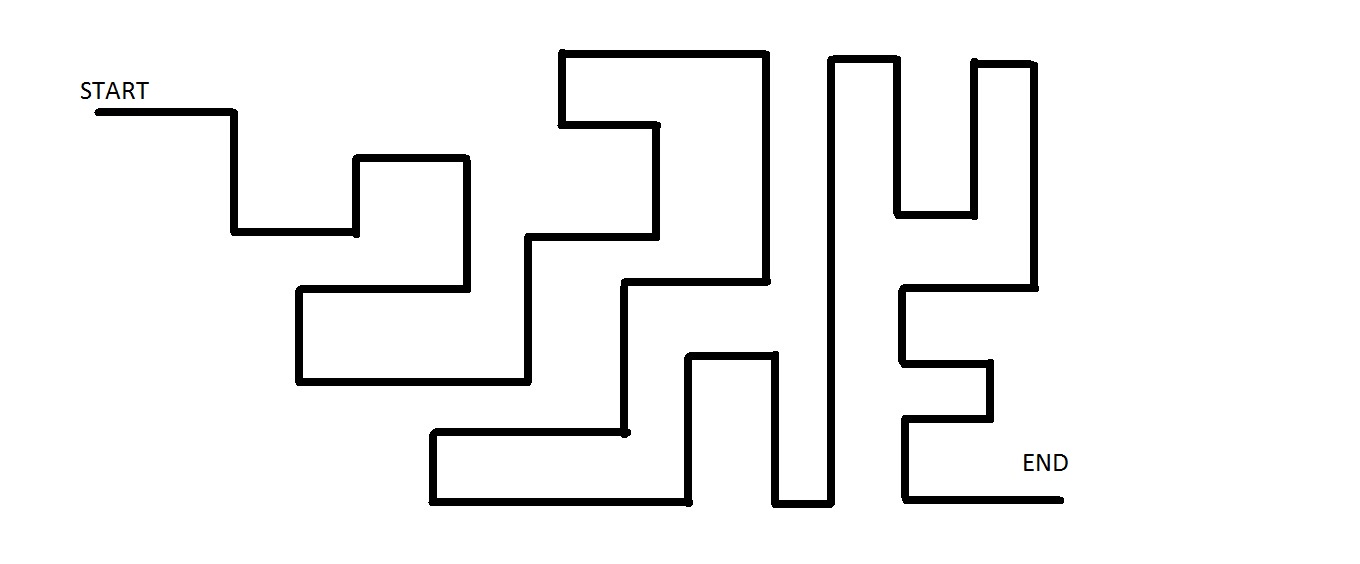}
\caption{TRACK}
\label{fig:track}
\end{figure*}

A line follower robot is an autonomous body expected to navigate in a network by following a specific line. The track on which the robot is expected to navigate is coloured black, and the background is white (figure \ref{fig:track}). Such devices are controlled by a person using a visual feedback from the scene of the track which is used to generate respective motor commands on the controller to control the robot. This involves multi sensory perception to compute the action required to be taken from the visual input and act accordingly by sending motor commands to the fingers to produce the desired effect on the robot navigating on the track. As this requires intervention in control from a human this can't be refereed to as a automated system.  The process is as shown in the figure \ref{fig:navghum}.

\begin{figure*}
\centering
\includegraphics[width=1\linewidth]{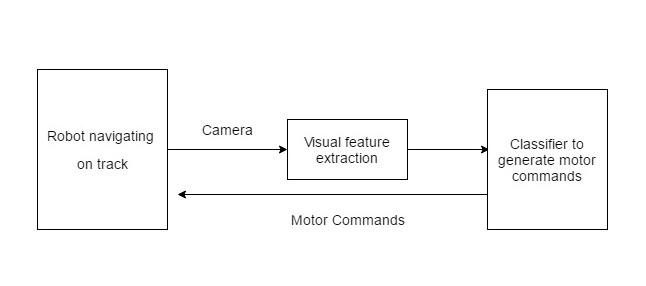}
\caption{Robot Navigation by Machine Learning classifier}
\label{fig:navgML}
\end{figure*}

 Several computer vision \cite{forsyth2002computer} \cite{umbaugh1997computer} along with machine leaning \cite{bishop2006pattern} paradigm has been successful to implement the same by making silicon chips to perform the computation that are to be performed by the brain. There are cameras on the robot which are performing the role of human eyes. The images from the camera gives some features regarding the path on which the robot is navigating. The direction on which the robot must move is classified based on the features extraction method and using a machine learning classifier. Once this is known, motor commands to make the robot move in desired direction is sent to the wheels. Hence the silicon chips and camera do the work of eye and multi sensory perception \cite{thakur2016dynamical}, of relating visual feedback to generate motor commands. Using computer vision techniques for extracting visual features followed by machine learning classifier to generate motor commands is shown in the figure \ref{fig:navgML}.\\
As specified the aim is to implement cell culture to make the controlling of robot automated.

\section{Cell cultures as controlling unit}

Multi electrode array - MEA \cite{obien2015revealing} recording gives a great opportunity to study network topology as it provides population recording. It provides data from multiple electrodes or channels which is thus effective for understanding network events. A MEA has culture on its surface from which electrical activity is recorded. A framework is modeled for using the electrical activity recording to control a robot to navigate following a line.   

Involvement of cell as a contriving unit will be done by following:
\begin{itemize}
\item By photo receptors cells
\item  By neural cells.
\end{itemize}

These two shall be dealt in following section.

\section{Photo receptor cells}
\begin{figure}
\centering
\includegraphics[width=\linewidth]{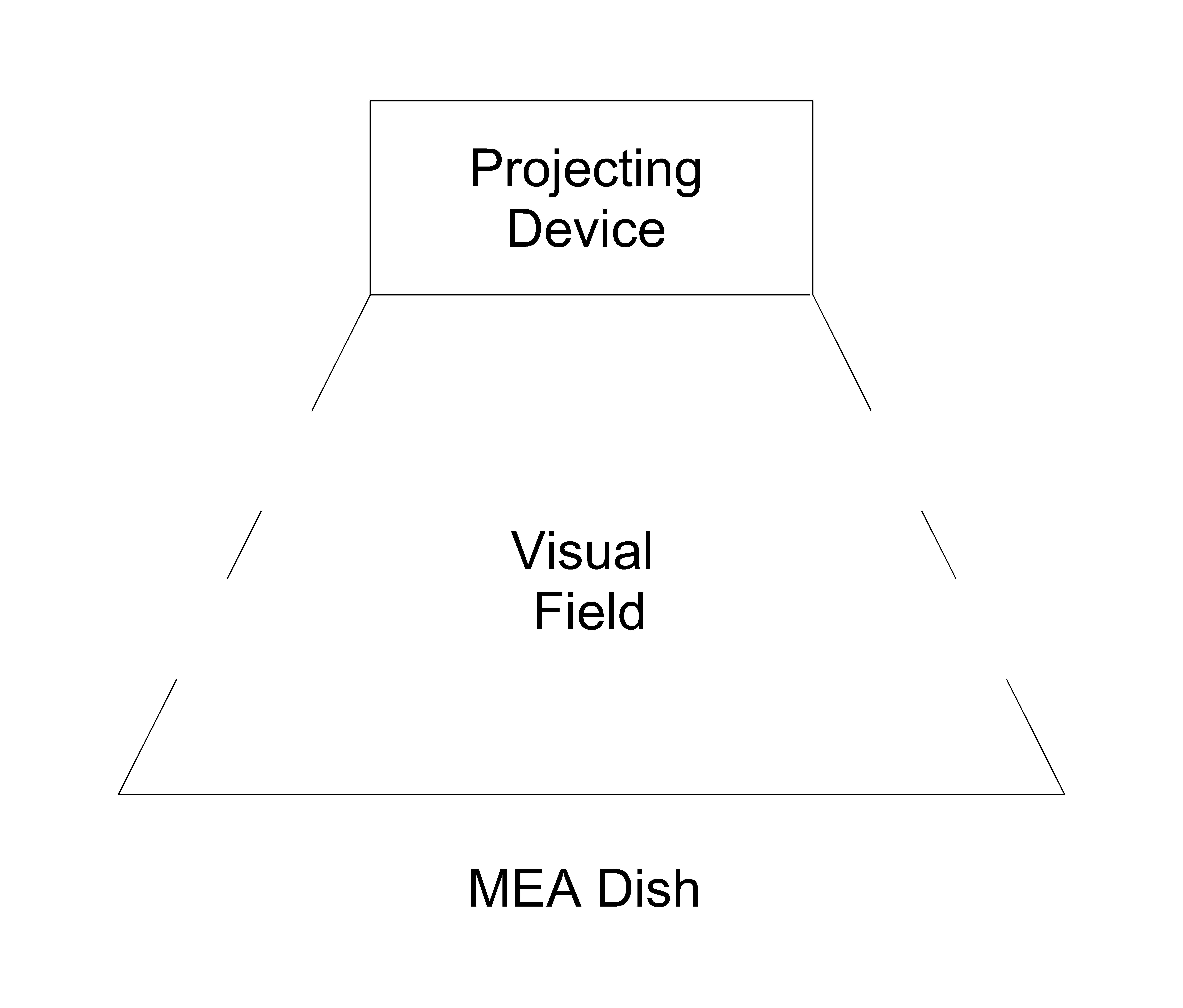}
\caption{Projection of the visual field captured through camera on top of the MEA dish containing photoreceptor cells}
\label{fig:proj}
\end{figure}

\begin{figure*}
    \centering
    \begin{subfigure}[b]{0.25\linewidth}        %% or \columnwidth
        \centering
        \includegraphics[width=\linewidth]{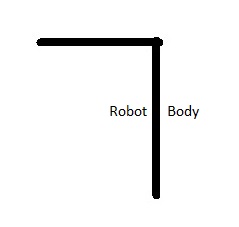}
        \caption{Visual field - A}
        \label{fig:lft}
    \end{subfigure}\quad
    \begin{subfigure}[b]{0.25\linewidth}        %% or \columnwidth
        \centering
        \includegraphics[width=\linewidth]{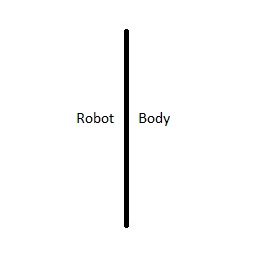}
        \caption{Visual field - B}
        \label{fig:str}
    \end{subfigure}
 \begin{subfigure}[b]{0.25\linewidth}        %% or \columnwidth
        \centering
        \includegraphics[width=\linewidth]{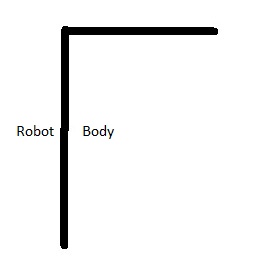}
        \caption{Visual field - C}
        \label{fig:rght}
    \end{subfigure}
    \caption{Visual field projected on MEA dish}
    \label{fig:vfield}
    \end{figure*}

Photo receptor cells are specialized cells in retina which performs photo transduction. Visual photo transduction is the method of converting light into electrical signals. Photo receptors are of great importance as they convert light into signals which stimulate biological processes. These cells absorb photons that causes a change in the cell's membrane potential. The model proposes usage of such cell culture on MEA dish for controlling a line follower. \\ The robot would be a simple system equipped with a camera to serve the purpose of visual feedback. As the camera takes up the current position of the robot, it is to be projected on the MEA dish containing photo receptor cell cultures. The usage of camera serves the purpose of eye. The projection of the visual field is shown in the figure \ref{fig:proj}. The track (figure \ref{fig:track}) if followed carefully one can find there are three possible different visual field as shown in figure \ref{fig:vfield}. The photo receptors respond by different electrical activity to different light intensities. Hence the projection of different visual field on the culture is expected to generate different electrical activity MEA dish. Hence to achieve control using photo receptor cell culture \cite{watanabe1990rod} it requires a mapping of electrical activity to visual field projected. This mapping would help to understand the visual field or current location of robot with the electrical activity recorded which would in turn help in generating motor commands. After the culture is done three of this possible visual field would be projected one by one, and corresponding electrical activity generated would be recorded. This will help in achieving the required mapping of visual field and electrical activity mapping.
Once the mapping is achieved the robot equipped with camera is left to be navigate on path. It would take visual field project it on the culture followed by which electrical activity shall be recorded based on which proper motor command would be generated and sent to the robot. \\
The methodology is as follows:
\begin{itemize}
\item Visual fields are projected on to the cell culture and mapping of electrical activity and visual field is obtained 
\item During test, the images are projected which will generate motor commands as per obtained mapping
\end{itemize}

\begin{figure}
\centering
\includegraphics[width=0.5\columnwidth]{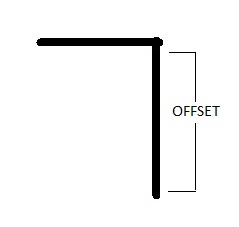}
\caption{Offset}
\label{fig:ofset}
\end{figure}

\begin{figure*}
\centering
\includegraphics[width=\linewidth]{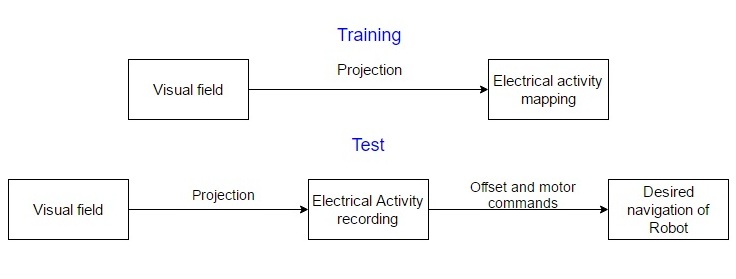}
\caption{Automated robot navigation by photo receptor cell}
\label{fig:navphrecep}
\end{figure*}

As per positioning of the camera in the experiment an offset has to determined. The offset would prevent the robot to take desired motor action as soon as visual field is detected. Offset is shown in figure \ref{fig:ofset}. If the robot would take say left turn on attending visual field A or electrical activity corresponding to visual field A, it would go off track out of line. This demands necessity of having a pre determined offset.
Hence motor commands required to be generated as per the visual field is given in table \ref{table:motcomm}. The procedure is illustrated in figure \ref{fig:navphrecep}.

\begin{table}
\caption{Motor Commands as per the Visual Field} % title of Table
\centering % used for centering table
\begin{tabular}{|c|c|} % centered columns (4 columns)
\hline\hline %inserts double horizontal lines
Visual Field & Motor Commands \\ [0.5ex] % inserts table
%heading
\hline % inserts single horizontal line
Visual Field A & Cover offset and take 90 $^{\circ}$ left \\ % inserting body of the table
Visual Field B & Go straight \\
Visual Field C & Cover offset and take 90 $^{\circ}$ right \\ \hline %inserts single line
\end{tabular}
\label{table:motcomm} % is used to refer this table in the text
\end{table}

\section{Neuron Cell}

\begin{figure*}
    \centering
    \begin{subfigure}[b]{0.25\linewidth}        %% or \columnwidth
        \centering
        \includegraphics[width=\linewidth]{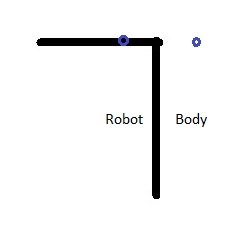}
        \caption{Sensor position - A}
        \label{fig:lftsns}
    \end{subfigure}\quad
    \begin{subfigure}[b]{0.25\linewidth}        %% or \columnwidth
        \centering
        \includegraphics[width=\linewidth]{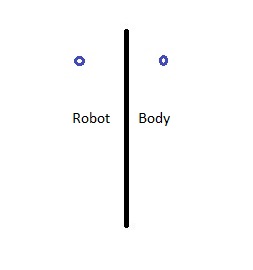}
        \caption{Sensor position - B}
        \label{fig:strsns}
    \end{subfigure}
 \begin{subfigure}[b]{0.25\linewidth}        %% or \columnwidth
        \centering
        \includegraphics[width=\linewidth]{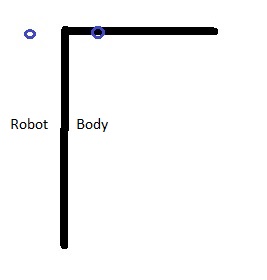}
        \caption{Sensor position - C}
        \label{fig:rghtsns}
    \end{subfigure}
    \caption{Sensor position of robot as it navigates. Blue circles denotes location of sensor.}
    \label{fig:snspos}
    \end{figure*}

\begin{figure*}
\centering
\includegraphics[width=\linewidth]{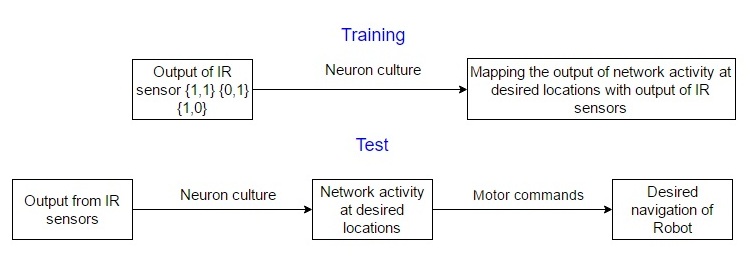}
\caption{Automated robot navigation by neural cell}
\label{fig:navneur}
\end{figure*}

\begin{figure}
\centering
\includegraphics[width=0.8\linewidth]{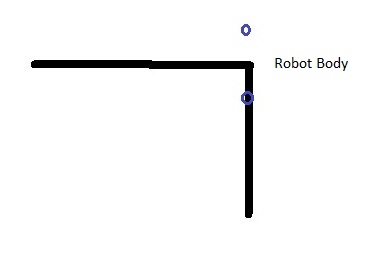}
\caption{Result of left turn from sensor position A - figure \ref{fig:lftsns}. Blue circles denotes location of sensor.}
\label{fig:wrngturn}
\end{figure}

Neuron cultures obtained generally from rat brain are cultured on MEA dish can be implemented as a control unit for making autonomous navigator. Framework that will be discussed here was proposed earlier \cite{george2015robot} for a autonomous system of navigating by avoiding obstacles. The proposed framework had a accuracy about 98\%.  It would be discussed how already existing framework could be utilized for modeling a line follower system. \\ The robot which shall navigate must have two infrared - IR sensors positioned pointing downward. As the robot navigates there could the only three possible states of view as shown in figure \ref{fig:snspos}. A IR sensor contains a emitter and detector. The emitter would emit infrared waves and purpose of detector would be to detect presence of any IR wave. As white reflects light and black absorbs it, same would be true with infrared wave. Hence if IR sensor detect (1) IR waves that means the colour is white and if it does not detect (0) the colour must be black.
Hence \{1,1\} \{0,1\} \{1,0\} corresponds to go straight, go left, go right respectively (figure \ref{fig:snspos} gives a pictorial representation). Hence knowing the output of IR sensor could predict the desired output motor command. Activity of MEA dish is used to understand the output of IR sensor. It was found that probability of spike occurrence at some particular electrode in response to stimulus sequence at some other electrode depends on electrode stimulation the order of stimulation and timing between the pulses \cite{george2014input}. A network of neuron is well able to distinguish between various inputs based on electrodes stimulated, the stimulating order and timing \cite{george2015robot}. This ability of neuronal network was used to encode inputs to a culture and decode output responses from the culture. From electrodes that excited the network to highest level stimulating electrode is chosen. Stimulating and recording electrodes are so chosen that by input pattern could be predicted from activity at recording electrodes \cite{george2015robot}. Hence the input pattern of  \{1,1\} \{0,1\} \{1,0\} could be predicted from activity patter of neuron cultures. As this prediction could be made foreseen motor commands could be generated making it possible for the robot navigate following a line.\\ Suppose the sensor position is as in figure \ref{fig:lftsns}. If then robot takes a 90 $^{\circ}$ left turn it results to a position described in figure \ref{fig:wrngturn} which is a undesired position. Similarly the system would attain a undesired position if robot takes a 90 $^{\circ}$ right turn from position C (figure \ref{fig:rghtsns}). This is undesired as one of the sensor is lying on white region where as the other is on black region. This issue could be resolved by generating a motor command to take a 90 $^{\circ}$ turn on the desired direction followed by a automated little forward push so that both sensor come into the white region. \\Hence automated navigator using neuron culture is developed. The corresponding motor commands are shown in table \ref{table:motcommsns}. The procedure is illustrated in figure \ref{fig:navneur}.

\begin{table}
\caption{Motor Commands as per the Sensor position} % title of Table
\centering % used for centering table
\begin{tabular}{|c|c|} % centered columns (4 columns)
\hline\hline %inserts double horizontal lines
Sensor Position & Motor Commands \\ [0.5ex] % inserts table
%heading
\hline % inserts single horizontal line
Sensor Position A & Take 90 $^{\circ}$ left and a little forward drift \\ % inserting body of the table
Sensor Position B & Go straight \\
Sensor Position C & Take 90 $^{\circ}$ right and a little forward drift \\ \hline %inserts single line
\end{tabular}
\label{table:motcommsns} % is used to refer this table in the text
\end{table}

\section{Discussion and Conclusions}
This was a implementation model using two type of cell culture in performing real world task of navigating following a line. It was shown how could the system be made autonomous by using two different type of cell cultures. The techniques for usage of neuron culture have been implemented practically and was found to have a accuracy of 98\% \cite{george2015robot}. It is expected that if the same methodology is extended in autonomous line follower system, as proposed, it would have a good accuracy too. The strong idea in this approach is using cell cultures as information processing tool, and to implement the ability for the purpose of classification of sensory inputs received and accordingly take decision of generating correct motor commands.
 Driving can be looked as a process of locomotion by obstacle avoidance and path following. A amalgamation of object avoiding paradigm \cite{george2015robot} and  line follower techniques, as suggested in thid work, could pave way for technological advancement of building new efficient and accurate autonomous driving systems.

\section{Acknowledgment}
Author would like to thank Shefali, Department of Food Technology and Biochemical Engineering, Jadavpur University, Kolkata for the valuable comments in improving the manuscript.
Author would also like to thank Department of Electrical Engineering, Jadavpur University, Kolkata. 
\balance

 \bibliographystyle{myIEEEtran} 
\bibliography{ref}

 \end{document}